\documentclass[preprint,authoryear]{elsarticle}
 
\usepackage{amssymb}
\usepackage{lipsum}
\usepackage{comment}
\usepackage{fullpage}
\usepackage{graphicx}	
\usepackage{amsmath}	
\usepackage{bm}
\usepackage{amssymb}	
\usepackage{wrapfig}
\usepackage{float}
\usepackage{adjustbox}
\usepackage{rotating}
\usepackage{booktabs}
\usepackage{ltablex} 
\usepackage{caption,tabularx}
\usepackage{longtable}
\usepackage{subcaption}
\usepackage{amsfonts}
\usepackage{multirow}
\usepackage{pdfpages}
\usepackage{hyperref}

\usepackage[acronym]{glossaries}

\usepackage{xcolor}
\usepackage[linesnumbered,ruled,vlined]{algorithm2e}
\NewDocumentCommand{\codeword}{v}{%
\texttt{\textcolor{black}{#1}}%
}

\journal{Astronomy and Computing}

\begin{document}

\begin{frontmatter}

\title{\texorpdfstring{$\beta$}{beta}-SGP: Scaled Gradient Projection with $\beta$-divergence for astronomical image restoration}

\author[1]{Yash Gondhalekar\corref{cor1}}
\ead{yashgondhalekar567@gmail.com}
\author[2]{Margarita Safonova}
\author[3]{Snehanshu Saha}

\cortext[cor1]{Corresponding author}

\address[1]{CSIS, BITS Pilani, K.K.~Birla Goa Campus, Goa, India}
\address[2]{Indian Institute of Astrophysics, Bangalore, India}
\address[3]{CSIS and APPCAIR, BITS Pilani, K.K.~Birla Goa Campus, Goa, India}

\begin{abstract}
Image restoration in astronomy has been considered a vital step in many ground-based observational programs that often suffer from sub-optimal seeing due to atmospheric turbulence, distortion of stellar shapes due to instrumental aberrations, trailing, and other issues. It holds importance for various tasks: improved astrometry, deblending of overlapping sources, faint source detection, and identification of point sources near bright extended objects, such as galaxies, to name a few. We conduct an empirical study by applying the Scaled Gradient Projection (SGP) iterative image deconvolution algorithm to restore distorted stellar shapes in our observed data. We investigate using a more flexible divergence measure, the $\beta$-divergence, which contains the commonly-used Kullback–Leibler (KL) divergence as a special case and allows automatic adaptation of the parameter $\beta$ to the data. An extensive set of experiments comparing the performance of SGP and its $\beta$-divergence variant ($\beta$-SGP) is carried out on extracted star stamps and on images containing multiple stars (both crowded and relatively sparser fields). We show a consistent enhancement in the flux conservation across all considered scenarios using $\beta$-SGP compared to SGP. Using a few quantifiable metrics such as the Full-Width-at-Half-Maximum (FWHM) and ellipticity of stars, we observe that $\beta$-SGP improves restoration quality, compared to the SGP, in many cases and still preserves restoration quality in others. We conclude that generalized versions of image restoration algorithms are more robust due to their enhanced flexibility and could be a promising modification for astronomical image restoration.
\end{abstract}
 
\begin{keyword}
methods: statistical \sep methods: numerical \sep techniques: image processing
\end{keyword}

\end{frontmatter} 

\section{Introduction}\label{introduction}

Image restoration in astronomy was considered to be a luxurious field until an ``impossible" mistake of spherical aberration was identified in the primary mirror of the Hubble Space Telescope (HST) in 1990 \citep{2001ISPM...18...11M}. Since then, much attention has been given, and considerable research has been carried out to develop novel techniques for deconvolving astronomical images. It is now widely believed that satisfactory deconvolution algorithms are crucial for maximizing the scientific information output from astronomical images, particularly for observations from ground-based telescopes that unpreventably suffer from blurring and distortion due to atmospheric seeing. As a result, the observed images are assumed to be convolved with the Point Spread Function (PSF), which quantifies intrinsic properties of the telescope, such as instrumental anomalies and the effects of atmospheric refraction, dispersion, and optical aberration due to the atmosphere. Some commonly faced issues in long-baseline astronomical observations include PSF anisotropy or smearing of starspots arising due to these reasons.

Due to the prevailing degradation of observed images in astronomy, it becomes important not to discard but restore them so they can be incorporated into further scientific analyses. For example, \citet{servilatt} discarded images with bad seeing or tracking problems from their observed dataset of the M13 globular cluster so that only good images remain. In cases where the observing program is over a long baseline, and the exposures are taken at a low cadence, each image becomes vitally important as it is impossible to repeat them (e.g., this issue is particularly important in the search for rare microlensing events in globular clusters since the data loss must be kept at the minimum to extract maximum information from them). Thus, in recent years, extensive efforts have been applied to develop deconvolution methods so that the full potential of large datasets can be utilized.

Deconvolution methods in astronomy have a long history, with works dating as early as the mid-1900s. A large amount of literature is devoted to the development of new image restoration/deconvolution methods with different focuses and applications; however, the primary aim of these methods is to correct or undo the effects of atmospheric and instrumental distortions (see, e.g., \citealt{2002PASP..114.1051S} and \citealt{2002SPIE.4847..144P} for a review of deconvolution methods). The Maximum Entropy Method (MEM) was applied in the image deconvolution context first by \citet{1984MNRAS.211..111S} and \citet{SkillingGull}, which maximizes the entropy of an image given some constraints (see \citealt{1986ARA&A..24..127N} for a review of MEM in astronomy) -- this formulation regularizes the ill-posed nature of deconvolution and thus tries to find the simplest solution possible. The CLEAN algorithm proposed by \citet{1974A&AS...15..417H} was another simple iterative procedure initially developed initially for mitigating issues due to irregular baselines in radio interferometry. \citet{1995ApJ...449..460K} proposed a mechanism for reversing the effects of PSF with a specific focus on weak gravitational lensing. Wiener filtering is another method but severely suffers from amplification of noise and ringing artifacts (\citealt{2002PASP..114.1051S}; \citealt{motionBlur}). The Richardson-Lucy (RL) algorithm \citep{1972JOSA...62...55R, 1974AJ.....79..745L}, also proposed in the 1900s, is a widely used deconvolution algorithm not only in astronomy but also in other disciplines. It has the benefit of ensuring the non-negativity of the image at each iteration. Perhaps, its wide use in the astronomical community is due to its simplicity in implementation, apart from the benefits it provides, and its broad generalizability to a variety of problems. However, several variants of RL need to be introduced to mitigate the enormous computing time requirements of RL due to its slow convergence (e.g., the multiplicative relaxation modification; \citealt{1986JOSAA...3..787M}).

Among these algorithms of great interest are the ones that not only reduce the distortion effects but also ensure flux conservation so that reliable photometry studies can be performed on the deconvolved images. For example, the RL algorithm under the assumption of zero background emission or the modification to the classical MEM, proposed by \citet{1996A&AS..118..575P}. Stating in general, deconvolution should facilitate scientific studies on the otherwise sub-optimal images. In this regard, the aim is not to make the deconvolved images nice-looking (although it does help for visualization), but the basic science requirements be met. With similar aims, the SGP algorithm, initially proposed in \citet{2009InvPr..25a5002B}, has proven promising for astronomical image deconvolution in recent years and as a possible improvement to the traditional RL algorithm. It was shown to yield computational benefits and potentially better reconstruction results than the standard Expectation-Maximization (EM) method \citep{EMmethod} and other related methods for astronomical image restoration (e.g., \citealt{2009InvPr..25a5002B}; \citealt{2012A&A...539A.133P}). SGP has been studied on astronomical sources ranging from point sources in open clusters to extended objects like the nebulae \citep{2012A&A...539A.133P}, or to restore motion-blurred star images obtained from a star sensor \citep{motionBlur}. SGP's convergence properties were discussed in previous works \citep{2015InvPr..31i5008B, convSGP2}, and further applications of SGP are described in \citep{2009InvPr..25d5010Z, bonneteniPrato, 2010InvPr..26b5004B}. Some recent applications of SGP include an SGP-based blind deconvolution approach studied in \citet{2017MNRAS.470.1950J} and an improved SGP method, along with a PSF estimation algorithm, proposed in \citet{iSGP}. More advanced studies on SGP's scaling strategies and projections onto generalized spaces have also been described in \citet{sgpScalingStudy} and \citet{sgpGeneralizedProjs}.

In this paper, we showcase the capabilities of the SGP algorithm for astronomical image restoration. Moreover, we test SGP with a flexible divergence measure, the $\beta$ divergence, as opposed to the commonly used KL divergence, and further allow the parameter $\beta$ to learn from the data (we call this approach $\beta$-SGP for ease of notation). The primary question we ask here is whether the flexibility allowed due to $\beta$-divergence shows any benefits for image restoration. Our experiments use simulated and real telescope images. The former case serves as a basic test to compare SGP and $\beta$-SGP since the conditions of the image are fully controlled. Our major focus, however, is to compare both algorithms on real observational data. Through a comprehensive empirical study of SGP and $\beta$-SGP,  using several quantifiable metrics, we show that due to its more flexible nature, $\beta$-SGP does show improved restoration results and exhibits better flux conservation than the SGP. We suggest that $\beta$-divergence is a promising alternative to the KL divergence for astronomical image restoration. For the majority of the paper, we have restricted our analyses to images containing stars, considered point sources.

\section{Methods}

\subsection{Dataset and Data Reduction}
\subsubsection{Simulated dataset}

We first test our modified SGP algorithm on two simulated images: $256 \times 256$ pixels HST image of the NGC 7027 planetary nebula and an image of a satellite. They have pixel values in the range $[0, 255]$. These images are distributed as part of the \texttt{SGP-dec} software available online\footnote{More information about \texttt{SGP-dec} and the simulated images can be found at \url{https://www.unife.it/prin/software}}. They have been artificially degraded by convolving the ground-truth images with a PSF, followed by the addition of a constant background level and perturbing the images with the Poisson noise. This allows us to compare the original \texttt{SGP-dec} implementation (in MATLAB) with our modified procedure. These data come with a PSF model, background level, the simulated ground-truth image, and the simulated degraded image to be deconvolved.

\subsubsection{Observational dataset}\label{sec:obsData}


To showcase our procedure's performance on real data, we test it on the image of Messier 13 (NGC 6205) globular cluster, henceforth M13. This image is part of a larger time-series dataset \citep{imbh}, obtained in 2008--2015 on the 2-m Himalayan Chandra Telescope (HCT) of the Indian Institute of Astrophysics (IIA), Leh, Ladakh, IAO, located at 4500 m above sea level. We selected the image, based on visual inspection, from the images with bad seeing and distorted stellar shapes, particularly to demonstrate the restoration algorithms' capabilities. The images are taken with the Himalayan Faint Object Spectrograph and Camera (HFOSC) mounted on the HCT. HFOSC is equipped with a Thompson CCD of $2048 \times 2048$ pixels with a pixel scale of $0''.296$/pix, equivalent to a total field of view (FOV) of $\sim$$10'\times10'$. The readout noise, gain, and readout time of the CCD are 4.87 $\bar{e}$, 1.22 $\bar{e}$/ADU, and 90 sec, respectively. The typical seeing values observed in the images from the dataset were $6-8.5$ pixels.

All images were subjected to the usual image reduction process (bias subtraction, flat-fielding, and cosmic rays removal) using IRAF (\citet{iraf1}; \citet{iraf2})\footnote{IRAF is distributed by the National Optical Astronomy Observatory, which is operated by the Association of Universities for Research in Astronomy, Inc., under a cooperative agreement with the National Science Foundation.} scripts. Flat fields were constructed from dithered images of the twilight sky, and any star images were removed by combining flats in each band using a median filter. Illumination correction was not required for our dataset because the detector's field of view is small, and the dark current is negligible since the CCD is cooled down to $-100^\circ$C. To remove the cosmic rays, we employed the IRAF task \codeword{crmedian}, which uses a median filtering approach to replace cosmic ray pixel values with the median value. We divide the full frame (2K$\times$2K pixels size) $I$-band image into $5 \times 5$ overlapping subdivisions, which we use for our study here. 

\subsection{Point Spread Function modeling}\label{sec:psfcalculation}

PSF models are good mathematical representations of the shapes of the stars. Synthetic PSF estimates that describe the PSF analytically are devoid of any degradations due to instrumental noise or other reasons, which could happen when the PSF is estimated from stars present in observational data (\citealt{2002PASP..114.1051S}; \citealt{2023arXiv230309422L} and refs therein). Moreover, deconvolution methods requiring the knowledge of PSF are strongly dependent on the quality of the PSF since unwanted artifacts can arise due to sub-optimal PSF models \citep{2023arXiv230309422L}. Hence, we have used the analytical approach by writing a custom code to model the PSFs.

The DIAPL package\footnote{We used the DIAPL code for Difference Image Analysis (DIA) \citep{Wozniak} as modified by Wojtek Pych. The package, along with its documentation, can be found at \url{https://users.camk.edu.pl/pych/DIAPL/}.} is an efficient implementation of the Optimal Image Subtraction (OIS) method described by \citet{1998ApJ} and \citet{Wozniak}. In practice, we found it efficient in handling overlapping stellar profiles in densely crowded fields like the globular cluster dataset considered in this paper. As a result, we use DIAPL to detect stars and use them to model the PSF.

DIAPL first finds candidate PSF stars in the input frame (using the \codeword{sfind} program) and then calculates the PSF model parameters, such as the raster size of the PSF model, $x$ and $y$ width scales of the PSF, number of Gaussian functions used to build the PSF model, degree of polynomial describing local and spatial dependent shapes, etc., using the \codeword{getpsf} program. DIAPL calculates the PSF model on the reference frame for image subtraction purposes. To cater it our purposes, we modified the \texttt{getpsf} program. We also divided the full frame into $5 \times 5$ subframes and estimated the PSF model in each subframe to eliminate any possible spatial dependence of the PSF. In \ref{appen-psfMatCalc}, we describe the procedure to convert a functional PSF model to a two-dimensional matrix representation, which would be one of the inputs to the image restoration algorithm.

\subsection{Deconvolution methods theory}\label{sec:deconvMethods}

The equation for an astronomical image acquired using a CCD can be given by: 
\begin{equation}
    \bm{g} = \bm{A} \bm{f} + \bm{b}\ + \bm{\eta}\,,
\end{equation}
where $\bm{g}$ is the observed degraded and noisy image from a telescope, $\bm{A}$ is the PSF, $\bm{f}$ denotes the undegraded noise-free image of the object that is unknown, and to be estimated, $\bm{b}$ denotes the background level, and $\bm{\eta}$ is the additive read-out noise. More specifically, modeling the observed image can be described by a mixed Poisson-Gaussian noise \citep{introInverseImaging}. In many practical cases, the additive read-out noise component is often modeled by Poisson noise (\citealt{1994rhis.conf..139S}; \citealt{cell2gal}; \citealt{interfImages}) or simply ignored for simplification (approaches using least-squares methods, that assume additive noise models, on images corrupted with Poisson noise also exist \citep{2005A&A...436..741V}), unless the images have very low photon count in which case taking into account the read-out noise might be important. We note that there have also been previous attempts to model both Poisson and Gaussian noise \citep[e.g., ][]{2005EJASP2005..103L, 2008InvPr..24c5016B, 1994rhis.conf..139S}.



It is a well-known fact that direct solutions to estimate $f$ is often unacceptable due to the ill-conditioned nature of the restoration problem \citep{2001ISPM...18...11M}. We also verify this by noting that the condition numbers of our modeled PSFs are much larger than one; they lie roughly in the range $10^{18} - 10^{27}$, which signifies that the matrices are ill-conditioned and near-singular. In such cases, its inverse calculation is prone to significant errors. It is suggested to use iterative restoration methods instead of finding direct inverse solutions (in a maximum likelihood setting, for example, we would like to move closer to the maxima of the likelihood function iteratively and use early-stopping (see Section~\ref{sec:terminationCriteria} for discussion on termination criteria)). Moreover, since such restoration approaches tend to be sensitive to noise, particularly when images have a low signal-to-noise ratio \citep{berryBurnell}, regularization is obtained by early stopping based on the number of iterations considering that one observes a semi-convergent behavior in the case where images are corrupted by noise \citep{introInverseImaging}. Deconvolution approaches, such as Wiener filtering, are not a good choice for such problems because they amplify image noise \citep{2002PASP..114.1051S, motionBlur} and inherently assume additive noise models, which is generally not a dominant source of noise in astronomical images \citep{1990ph...book.....L}.

\subsection{SGP and its generalized version using \texorpdfstring{$\beta$}{beta}-divergence}\label{sec:sgp-restoration}
The SGP algorithm, originally proposed in \citet{2009InvPr..25a5002B}, is also termed as an efficient version of the famous RL deconvolution algorithm (\citealt{1972JOSA...62...55R}; \citealt{1974AJ.....79..745L}) that allows for faster convergence via effective scaling and step length strategies and potentially better reconstruction results than RL. It is an iterative procedure that minimizes an objective function with some desirable properties, along with at least a non-negativity constraint on the reconstructed image. Specifically, SGP tries to solve (under the non-negativity and flux conservation constraints):
\begin{equation}
\begin{split}
    \textrm{min} \,\, & J(\bm{f}; \bm{g})\\
    \textrm{sub. to} \,\, & \bm{x} \geq \bm{0}, \quad \sum_{i=1}^{n} x_{i} = c, \medspace i \in S\,,
\end{split}
\label{eqn:klDivEquation}
\end{equation}
where $J$ is a continuously differentiable function (although the case of convex objective functions was dealt with in the original paper), $\bm{f}$ is the restored image at some iteration, $\bm{g}$ is the observed image, $S$ is the source region, and $c$ is some constant. Under the Poisson noise assumption (a commonly used model for astronomical images), the function $J$ in Eq.~\ref{eqn:klDivEquation} can be chosen to be the KL divergence of the blurred and background-added version of the partially reconstructed image ($A\bm{f} + b$) from the observed image $\bm{g}$. KL divergence is the traditional divergence measure used in SGP and has great advantages since minimizing the KL divergence is equivalent to maximizing the likelihood function, and as such, it is theoretically justified.

Here, our aim is to explore a generalized divergence measure, the $\beta$-divergence (\citealt{basu1998}, \citealt{eguchi2001}), which can be described as:
\begin{equation}
    \begin{split}
        d_{\beta}(\bm{g}, \bm{f})=\begin{cases}
        \dfrac{1}{\beta (\beta - 1)}\left(\bm{g}^{\beta} + (\beta - 1) \bm{f}^{\beta} - \beta \bm{g}\bm{f}^{\beta - 1}\right)\,, \\ & \hspace{-0.42in} \text{if $\beta \in \mathbb{R} \backslash \{0, 1\}$}\,.\\
        \bm{g}\log{\dfrac{\bm{g}}{\bm{f}}} - \bm{g} + \bm{f}, & \text{if $\beta = 1$}\,.\\
        \dfrac{\bm{g}}{\bm{f}} - \log{\dfrac{\bm{g}}{\bm{f}} - 1}, & \text{if $\beta = 0$}\,.
      \end{cases}
    \end{split}
\end{equation}
Note that while the authors in the above references considered the case $\beta \geq 1$, the definition can be extended for all real values of $\beta$, as described in \citet{fevotte}. It is a family of functions parametrized only by a single parameter $\beta$ that controls the trade-off between the robustness and efficiency of the estimators of parameters. The special cases of $\beta = 2, 1, 0$ correspond to the Euclidean distance, the generalized KL divergence, and the Itakura-Saito divergence, respectively. Specific values of $\beta$ can be used if one has complete knowledge of the noise model: $\beta = 2$ can be used in cases of Gaussian noise, $\beta = 1$ for Poisson noise, and $\beta = 0$ for multiplicative Gamma noise \citep{cemgilfevotte}. A neat feature of this class of divergence is that it smoothly connects commonly-known distance measures described above. A common application of $\beta$-divergence has been in non-negative matrix factorization (\citealt{kompass}; \citealt{fevotte}). Another peculiar feature of the class of $\beta$-divergences is the robustness to outliers which was exploited in various applications (\citealt{basu1998}; \citealt{rbs}; \citealt{rvae}).

As stated above, astronomical images acquired from CCD consist of a combination of Gaussian and Poisson noises, and hence a Poisson noise assumption is not perfect. Such an assumption is valid in many cases (e.g., by approximating Gaussian noise by Poisson noise). Moreover, it has been argued that modeling both noise models separately is only useful in certain scenarios \citep{cell2gal}. However, we hypothesize that a value of $\beta \in (1, 2]$, or some value around $\beta = 1$, could serve as a better loss function due to the complicated nature of noise in real observational images, and verify whether it is helpful in different scenarios. We also expect such a modification to work better in simulated cases where multiple noise models are used, e.g., a combination of Gaussian and Poisson noise.

While not necessarily required to exploit the flexibility due to $\beta$-divergence, we exploit the parameterized nature of $\beta$-divergence by adapting $\beta$ to the data using a stochastic gradient descent procedure. This is to investigate whether a suitable $\beta$, learned from the data, can outperform the theoretically-rationalized choice of KL divergence. A fair expectation could be that the adaptation of $\beta$ to the data should make the starting choice of $\beta$ less important since the updates should naturally move towards the optimal $\beta$, if any. However, this may not always be the case since the loss function takes a new form parametrized by $\beta$ for every updated $\beta$. The recipes needed are: calculating the $\beta$-divergence of the blurred and background-added version of the reconstructed version from the original image, its derivative, which is defined as:
\begin{equation}
    \nabla J(\bm{f}; \bm{g}) = (A\bm{f} + \bm{b})^{\beta - 1} - A^T \bm{g} (A\bm{f} + \bm{b})^{\beta - 2}\,,
\end{equation}
and the derivative of $\beta$-divergence with respect to (w.r.t) $\beta$, for updating the parameter $\beta$. Only the last recipe is a new one (and only needed if one seeks to update $\beta$ as iterations proceed), and the other two are simply the $\beta$-divergence equivalent of KL divergence as used in the original SGP algorithm. Thus, updates to $\beta$ are regulated by the gradient of $\beta$-divergence w.r.t $\beta$, which contains information on which direction $\beta$ should move to. It is important to note that the $\beta$-divergence is convex w.r.t $A\bm{f} + \bm{b}$ for $\beta \in [1, 2]$ \citep{fevotte}, so one can restrict $\beta$ within this range if convexity is desired. However, we here perform no such restriction. We also note that adapting $\beta$, as done here, is not required, in which case one must appropriately fix $\beta$. Further details of the SGP algorithm, including the modification to update the parameter $\beta$ and a few other changes, are described in~\ref{appn:sgp-details}.


\subsubsection{Termination criterion}\label{sec:terminationCriteria}

Several stopping rules can be used, depending on the problem, to prevent amplification of noise during the iterations (see Section~3.3 of \citet{2012A&A...539A.133P} for a review of some rules). For point sources in simulation studies, SGP can generally be pushed to a convergence \citep{2012A&A...539A.133P} (see \citealt{lbt} for a similar observation in the case of the RL algorithm).

Here, we terminate based on the convergence of the objective (or the data fidelity) function such that the iteration is stopped when
\begin{equation}
    |J(\bm{f}^{k+1}; \bm{g}) - J(\bm{f}^{k}; \bm{g})| \leq tol \enspace J(\bm{f}^{k}; \bm{g})\,,
\end{equation}
where $tol$ is the tolerance level which we set to $10^{-4}$ unless otherwise specified. Apart from this, we also set a maximum iteration limit of 500 for experiments on non-simulated images, which provides regularization.

\section{EXPERIMENTS AND RESULTS}\label{sec:expRes}
Here we compare the use of $\beta$-divergence in SGP with the original SGP algorithm via experiments: on two simulated images, on individual star stamps extracted from observed globular cluster images, and on larger-sized subdivisions extracted from observed globular cluster images containing multiple stars in a single image.

\subsection{Setup and evaluation metrics}\label{sec:setup}

We demonstrate the capabilities of the SGP algorithm and leverage statistical metrics to compare SGP and $\beta$-SGP. All experiments were conducted using Python 3.8.10 on a computer equipped with Intel(R) Core(TM) i3-1005G1 CPU processor at 1.20 GHz.



It is known that even if SGP has a lot more parameters than RL, extensive experiments have led to an optimization of the parameters such that no specific parameter tuning is needed irrespective of the application \citep{2012A&A...539A.133P}. For this reason, we do not perform any SGP parameter tuning and use the default values described earlier (\citealt{2012A&A...539A.133P} noted that any choice of the step length parameter $\alpha_{k}$ belong to $[\alpha_{min}, \alpha_{max}]$ is a valid choice, and that $\alpha$ can be tuned inside this interval to optimize the reconstruction performance; however, we have not experimented with different values of $\alpha$). It has been found that including the flux constraint does not remarkably improve the convergence rate of SGP (see Fig.~4 in \citealt{2009InvPr..25a5002B}). However, due to photometry considerations, we have necessarily used the flux conservation constraint. It is also important to note that our implementation here does not include boundary effect correction (see \citealt{bertero2005} and \citealt{bertero2013} for discussion on this issue).

For all the experiments using $\beta$-divergence, we adapt the $\beta$ parameter using a simple gradient descent procedure, as described in Sect.~\ref{sec:sgp-restoration}. We have also allowed the learning rate to be adaptive using an exponential decay schedule, which is described further in Sect~\ref{appn:sgp-details}.

\subsection{Comparison on simulated images}\label{sec:simResults}
We first perform a test on a simulated image of the NGC7027 nebula and an image of a satellite that came as part of the test cases in the \texttt{SGP-dec} software. Using these two examples, we aim to compare the traditional SGP with the $\beta$-SGP we introduce in this paper. Since the ground-truth image is available in simulations, we use the relative L2 error as the metric, as described in the tests in \texttt{SGP-dec}. The same initialization of the restored image and the stopping criterion for both approaches are used and are taken from the \texttt{SGP-dec} software tests: a constant image with pixel values equal to a normalized flux value, and stopping the iterations is a fixed number of iterations. For SGP, 27 and 332 iterations, respectively, were considered optimal, and we used the same values for both SGP and $\beta$-SGP.

For selecting the starting value of $\beta$ (called as $\beta_{init}$) for both cases, we randomly sample 30 different $\beta$ values from a normal distribution with a mean of 1 and a standard deviation of 0.05 for NGC7027 and 0.01 for the satellite image. Such a selection procedure uses the fact that it is expected that the optimal starting value of $\beta$ is not too far away from 1 (corresponding to KL divergence). It was also noted in \citet{basu1998} that larger $\beta$ values were found to be less efficient. We then select that $\beta_{init}$ that yields the lowest relative error. The final run is made with this $\beta_{init}$.

Table~\ref{table:simResults} shows the relative L2 error achieved at the iteration numbers 27 and 332 and the starting value of the parameter $\beta$ in $\beta$-divergence that gives the lowest error. It shows that the optimal starting point is close to 1, as expected. In particular, a decrease in the minimum relative error is observed when using $\beta$-divergence as compared to using KL divergence. These two cases show that a $\beta$ value different from 1 was able to yield a lower error than fixing it to $\beta = 1$. This is likely due to the enhanced flexibility of $\beta$-divergence, and thus it was able to ``fit" the data better.

\begin{table}
       \centering
       \caption{Performance comparison of the original SGP and SGP with $\beta$-divergence ($\beta$-SGP) in terms of the minimum relative error achieved. Both approaches use the same number of iterations.}
       \begin{tabular}{ccccc}
        \hline
        Image & SGP min. rel. error\footnote{Slight differences in the minimum relative L2 errors can be observed between the original MATLAB SGP implementation and our Python implementation.} & $\beta$-SGP min. rel. error & $\beta_{init}$\\
        \hline
        NGC7027 & 0.1379 & 0.1366 & $\sim$0.9887\\
        Satellite & 0.2932 & 0.2912 & 1.0001\\
        \hline
       \end{tabular}
       \label{table:simResults}
   \end{table}


\subsection{Comparison on star stamps}\label{sec:starStampsResults}
We now compare results on square-sized star cutouts extracted from the observational dataset of the M13 globular cluster, as detailed in Sect.~\ref{sec:obsData}. Here the aim is to check the performance on cutouts containing a single star, which will serve as a first test. As a result, we have selected stars from the outskirts of the cluster to prevent any potential blending of stars. For detecting the stars used in this experiment, we have used the DIAPL package's \texttt{sfind} routine. It can be substituted with any other software one chooses since the further procedure is agnostic to the software choice. To extract star cutouts, we leverage the \texttt{photutils} package's \texttt{Cutout2D} class. The cutout size can be set such that the cutout serves as a minimal bounding box surrounding the star, padded with a few pixels to ensure full containment of the star's wings. For background estimation, we use the \texttt{Background2D} class from \texttt{photutils} and use the median value of the background map to yield a scalar background level. We use the PSF model of the subframe from which the star is extracted. The flux is estimated as $\sum I_{i} - N * bkg$, where $N$ is the total number of pixels in the observed star cutout, $I$, with a background level, $bkg$. For all the experiments here and henceforth, we use the observed (unrestored) image as the initialization to the SGP and $\beta$-SGP algorithms.


We randomly selected $\sim$1100 star cutouts extracted from all time-series M13 images in the dataset. For each cutout, we run SGP and $\beta$-SGP and compare their performance. Similar to Sect.~\ref{sec:simResults}, we need to select an optimal starting $\beta$ value. Hence, we use five trial $\beta$ values sampled from a normal distribution with a mean of 1 and a standard deviation of 0.05 and select that $\beta_{init}$ that gives the best flux conservation, i.e., with the least flux difference between the restored and the original star cutout. In principle, $\beta_{init}$ can be optimized using a combination of one or metrics, but we have only used flux conservation for simplicity.


We first show the results across all the cutouts used in this experiment using the FWHM, ellipticity, and flux conservation as the metrics for quantifying the performance (as described in Sect.~\ref{sec:setup}). For calculating the metrics to quantify restoration quality, we have used the \texttt{SourceCatalog} class from \texttt{photutils} \citep{astropyPhotutils} to yield the FWHM, ellipticity, and flux. For detecting sources, we use the \texttt{SourceFinder} class and use deblending. We also set the \texttt{npixels} parameter to 5 for the original image and 1 for the restored images since the restored stars are expected to be much more compact than the original ones. All other options are kept as default, as of \texttt{photutils v1.5.0}.

\begin{figure*}
  \begin{subfigure}[t]{.32\textwidth}
    \centering
    \includegraphics[keepaspectratio,width=\linewidth]{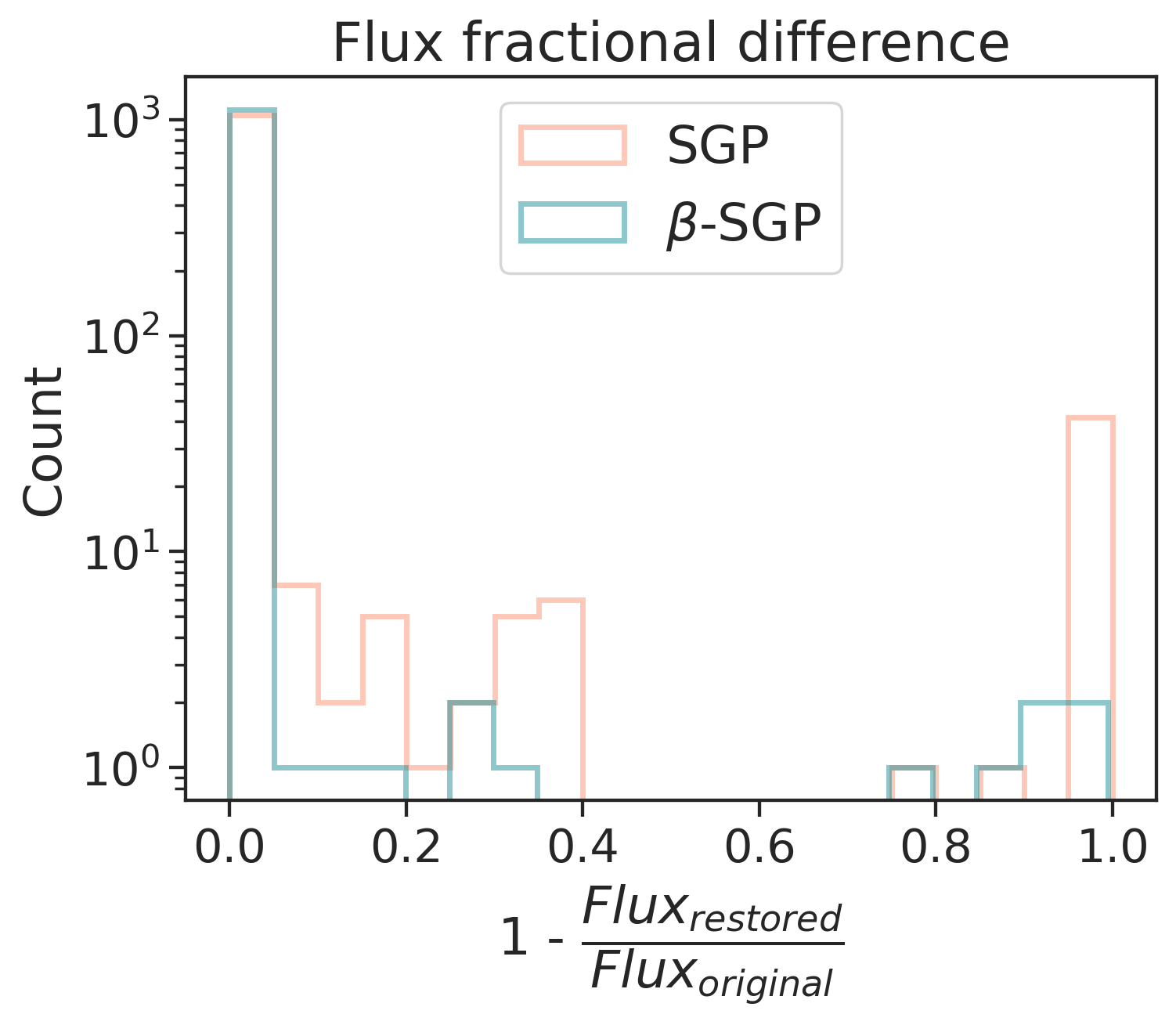}
    \caption{Mean flux fractional difference for SGP and $\beta$-SGP = 0.044 and 0.006, respectively. The median flux fractional difference is $\sim$0 for both.}
    \label{subfig:StampsFlux}
  \end{subfigure}
  \hspace{1em}
  \begin{subfigure}[t]{.32\textwidth}
    \centering
    \includegraphics[keepaspectratio,width=\linewidth]{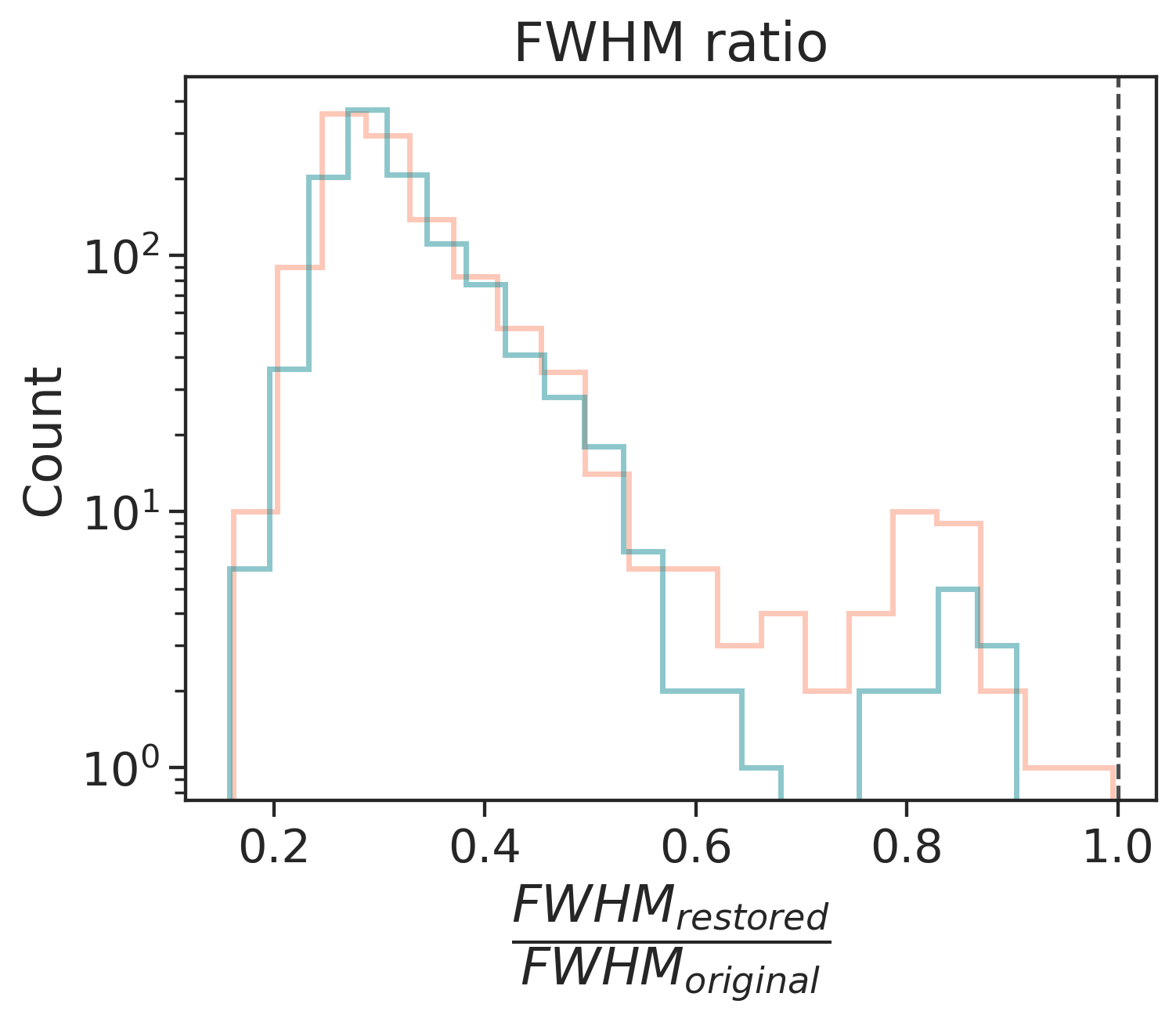}
    \caption{Mean and median ratio for SGP and $\beta$-SGP = (0.33, 0.32), and (0.3, 0.3), respectively}
    \label{subfig:stampsFWHM}
  \end{subfigure}
  \hspace{1em}
  \begin{subfigure}[t]{.32\textwidth}
    \centering
    \includegraphics[keepaspectratio,width=\linewidth]{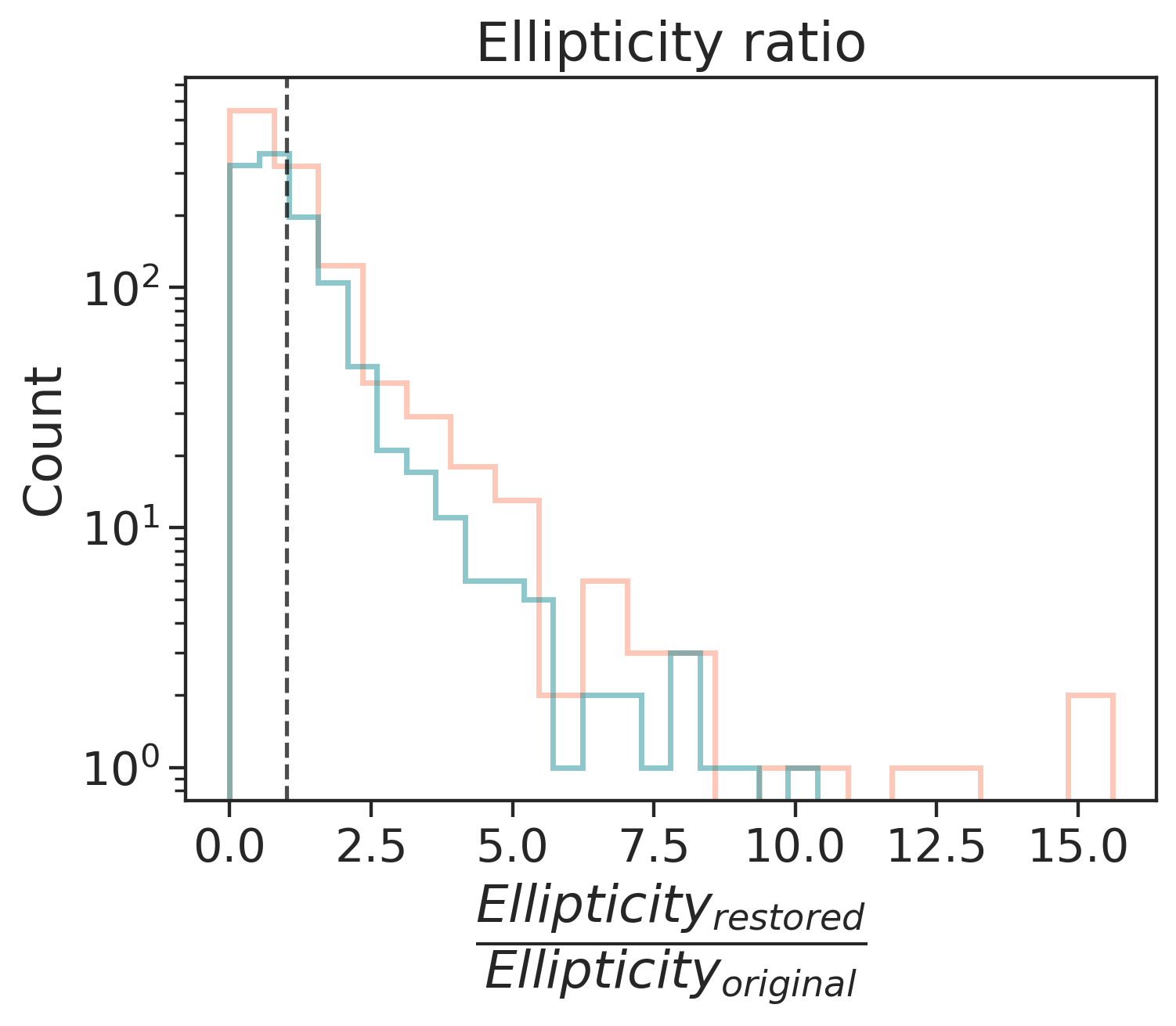}
    \caption{Mean and median ratio for SGP and $\beta$-SGP = (1.19, 1.12), and (0.79, 0.83), respectively}
    \label{subfig:stampsEll}
  \end{subfigure}
  \caption{Statistical comparison of SGP and $\beta$-SGP's restoration of $\sim$1100 observed star stamps using three metrics: flux, FWHM, and ellipticity. The vertical dashed black line in (b) and (c) denotes a ratio of 1.}
  \label{fig:comparisonStamps}
\end{figure*}

Fig.~\ref{fig:comparisonStamps} shows a histogram comparison of both approaches using these three metrics. In Fig.~\ref{subfig:StampsFlux}, for satisfactory flux conservation, the fractional difference must be close to zero. It shows that $\beta$-SGP preserves the flux better since it contains more examples where the fractional difference in flux is close to zero. Statistically speaking, exact flux conservation is not guaranteed. Hence, we must set an error bar around the original star's flux such that any restored flux lying within that error bar will be considered satisfactory flux conservation. For example, using 0.1\% times the original flux as the error bar, $\beta$-SGP conserves flux satisfactorily in $\sim$99\% cases, whereas SGP in $\sim$93\% cases. One can observe that no fractional difference value is $> 1$, which indicates that flux is never overestimated by both SGP and $\beta$-SGP but can be underestimated.

We next compare the FWHM and ellipticity ratio of the restored and original star under the two approaches. The aim is to see whether SGP and $\beta$-SGP make the stars more compact (since the blurring effect by PSF is reversed) and reduce their ellipticity (so that the PSFs are circularized and hence become closer to an ideal PSF; see Fig.~\ref{fig:1} for a visual demonstration). Hence, the lower the ratio, the better. In Fig.~\ref{subfig:stampsFWHM} and \ref{subfig:stampsEll}, the majority (in fact, all for FWHM) of the ratios are $< 1$, which indicates that SGP and $\beta$-SGP both improve the star's shape overall. However, there are a non-trivial number of cases for SGP and $\beta$-SGP where ellipticity increases. In terms of these two metrics, $\beta$-SGP and SGP show similar performance overall based on the mean/median FWHM and ellipticity ratio (see captions of Fig.~\ref{subfig:stampsFWHM} and \ref{subfig:stampsEll}). However, it can be seen that $\beta$-SGP has fewer instances of an FWHM/ellipticity ratio closer to one compared to SGP, which shows that $\beta$-SGP has a slight advantage in terms of these two metrics.


\begin{figure}
    \centering
    \includegraphics[width=\textwidth,keepaspectratio]{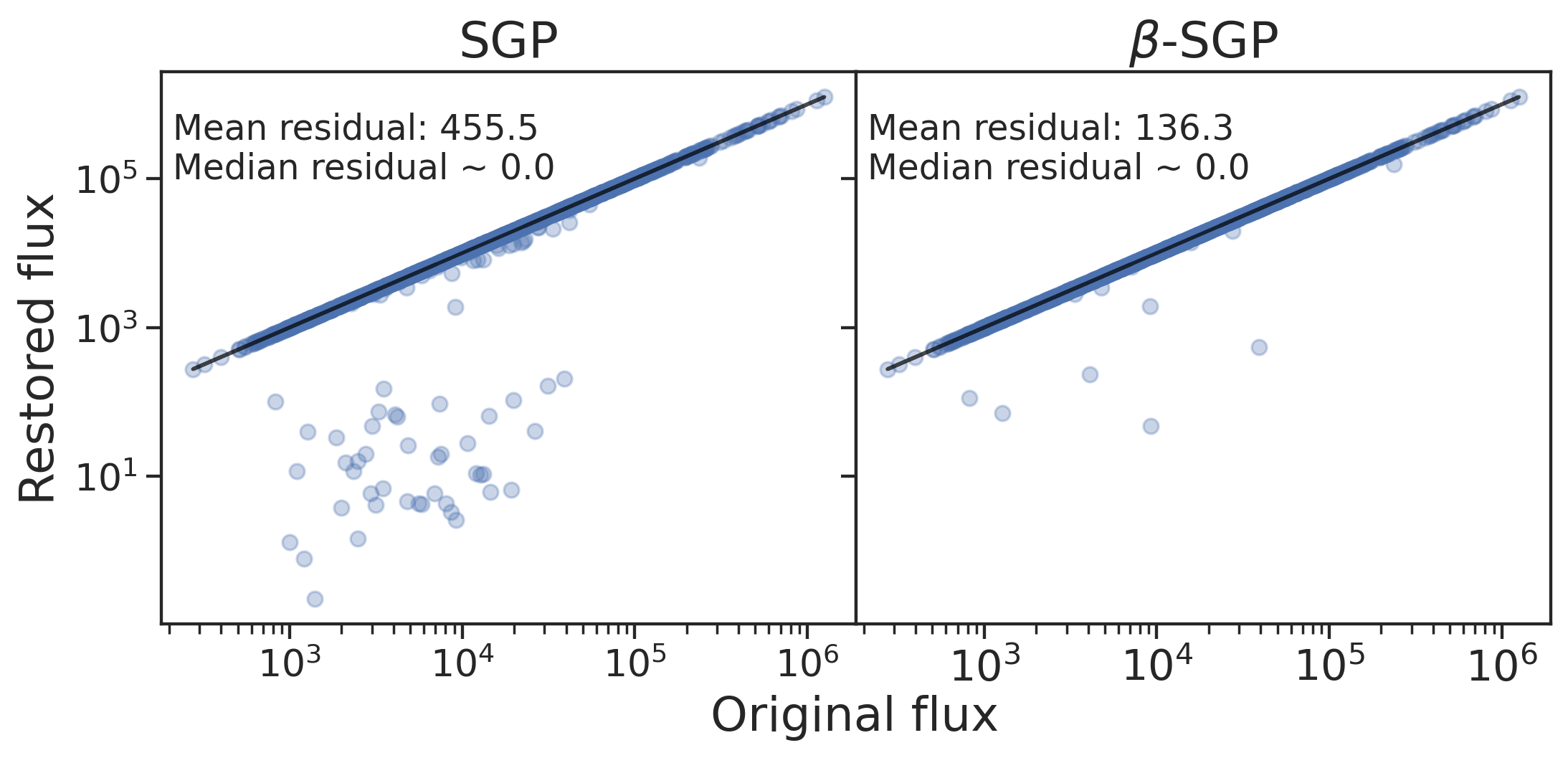}
    \caption{Comparison of original and restored fluxes (in counts) of stars from all the extracted star stamps. The blue points represent the actual data points, whereas the black line denotes perfect flux conservation. Note that these residual measures are different in spirit to the flux fractional difference shown in Fig.~\ref{subfig:StampsFlux}}
    \label{fig:fluxLinePlotStamps}
\end{figure}

An important consideration is how the flux conservation performance changes as the flux of the original source changes. This is because it has been well-known that deconvolution performance becomes increasingly unhelpful as sources become fainter (e.g., \citealt{1992MNRAS.259..104T}). As a result, we compare the original and restored flux for a star-by-star comparison. In Fig.~\ref{fig:fluxLinePlotStamps}, we plot the original vs. the restored flux using SGP and $\beta$-SGP. It is clear that the flux is conserved well for all ranges of original fluxes (as low as $\sim 5 \times 10^2$ to $10^6$) for both SGP and $\beta$-SGP. However, SGP shows non-trivial differences between the original and restored fluxes in the mid-flux range ($\sim 5 \times 10^3 - 5 \times 10^4$), which is absent in $\beta$-SGP, which shows that $\beta$-SGP preserves the flux better. There are also significant deviations from the straight line for SGP and $\beta$-SGP -- these are outlier cases and are again lesser in number for $\beta$-SGP than SGP. Moreover, these large deviations occur mainly for smaller original fluxes ($\lesssim$ $10^4$). $\beta$-SGP's superior flux conservation is quantified by the mean residuals shown on the plot. It is not unexpected to find the median residual to be zero and the same for both approaches since the no. of points with significant deviation from perfect flux conservation is small compared to the total number of points.

We have found no significant evidence supporting the idea of whether the FWHM and ellipticity ratios $> 1$ mainly correspond to lower flux (i.e., fainter) stars in the original images for SGP and $\beta$-SGP. This is because only $\sim$15-20\% stars in the original image, with flux less than 20-quantiles of the entire flux distribution, have an ellipticity ratio $> 1$. Note that there is no case where the FWHM ratio is $> 1$.

\subsection{Comparison on fields containing multiple stars}
The previous section served as a basic test using a conservative scenario of isolated stars. For practical purposes, several complications arise. It has been observed that while the restoration algorithms can be applied to star-field images containing multiple stars from simulated images, their application on real star-field images is not straightforward, in that it tends to suppress the dimmer stars and enhances the brighter stars \citep{minEntropyRestoration}. In SGP, it is speculated that the projection onto the non-negative orthant could be responsible for the suppression of the fainter stars for which \citet{2012A&A...539A.133P} suggested a possible remedy: to decrease the tolerance used in the KL divergence stopping criterion in SGP. For all experiments in this section, we set the tolerance, $tol = 10^{-5}$. On top of that, to prevent suppressing fainter stars, for detecting sources, we have used a threshold of 1.5$\sigma$ so that the relatively fainter stars are detected, and hence their flux is considered when considering SGP and $\beta$-SGP's flux constraint. We speculate that using larger thresholds such as 3$\sigma$ can disregard flux contribution from any fainter but true sources, which is undesirable. This becomes increasingly important in fields where many detectable, relatively faint sources exist apart from bright ones.

\subsubsection{Subdivision extracted away from the center}
\begin{figure}
    \centering
    \begin{subfigure}{\textwidth}
        \includegraphics[width=\textwidth,keepaspectratio]{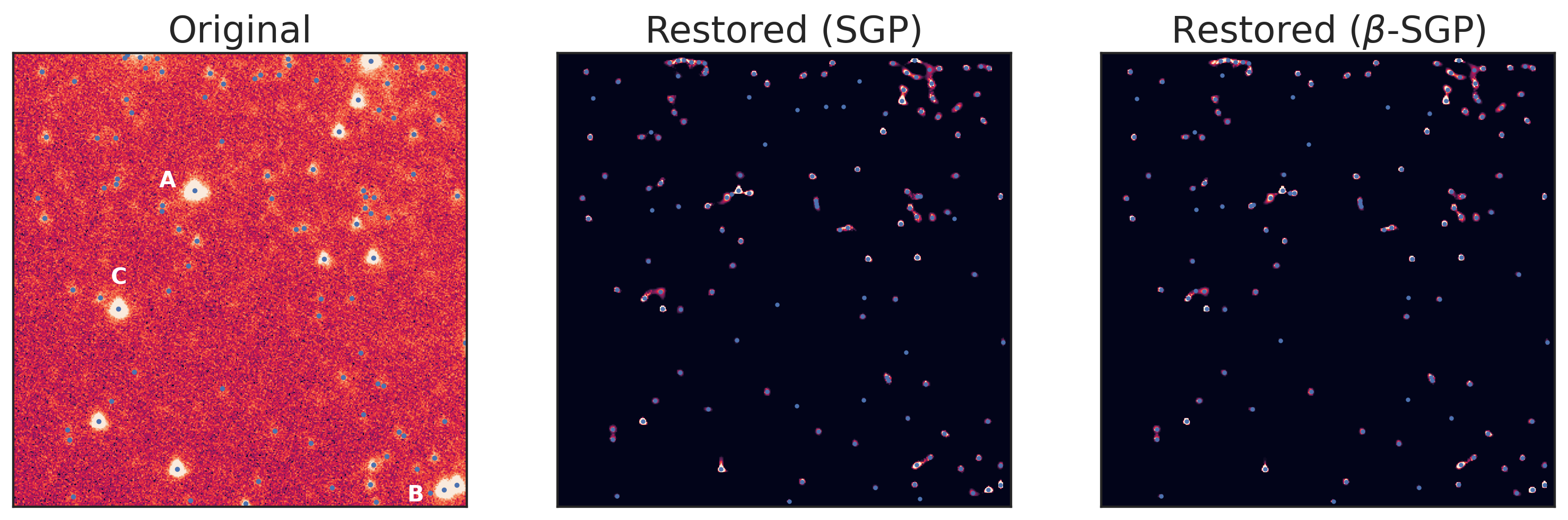}
        \caption{Detected sources are marked with a blue dot. The original image contains 103 detected sources, and the restored image using SGP and $\beta$-SGP contain 134 and 126 sources, respectively. Three positions are marked as A, B, and C, discussed further in the text. SGP and $\beta$-SGP required 51 and 43 iterations, respectively, and execution times of $\sim$6.5s and $\sim$6.7s, respectively. The optimal $\beta_{init}$ for $\beta$-SGP found was $\sim$0.97. Images are shown using a combination of square root stretching and clipping values outside the central 98\% percentile pixel values for the original and 99\% for both the restored images.} \label{fig:ex1}
    \end{subfigure}%
    \hfill
    \begin{subfigure}{\textwidth}
        \includegraphics[width=\textwidth,keepaspectratio]{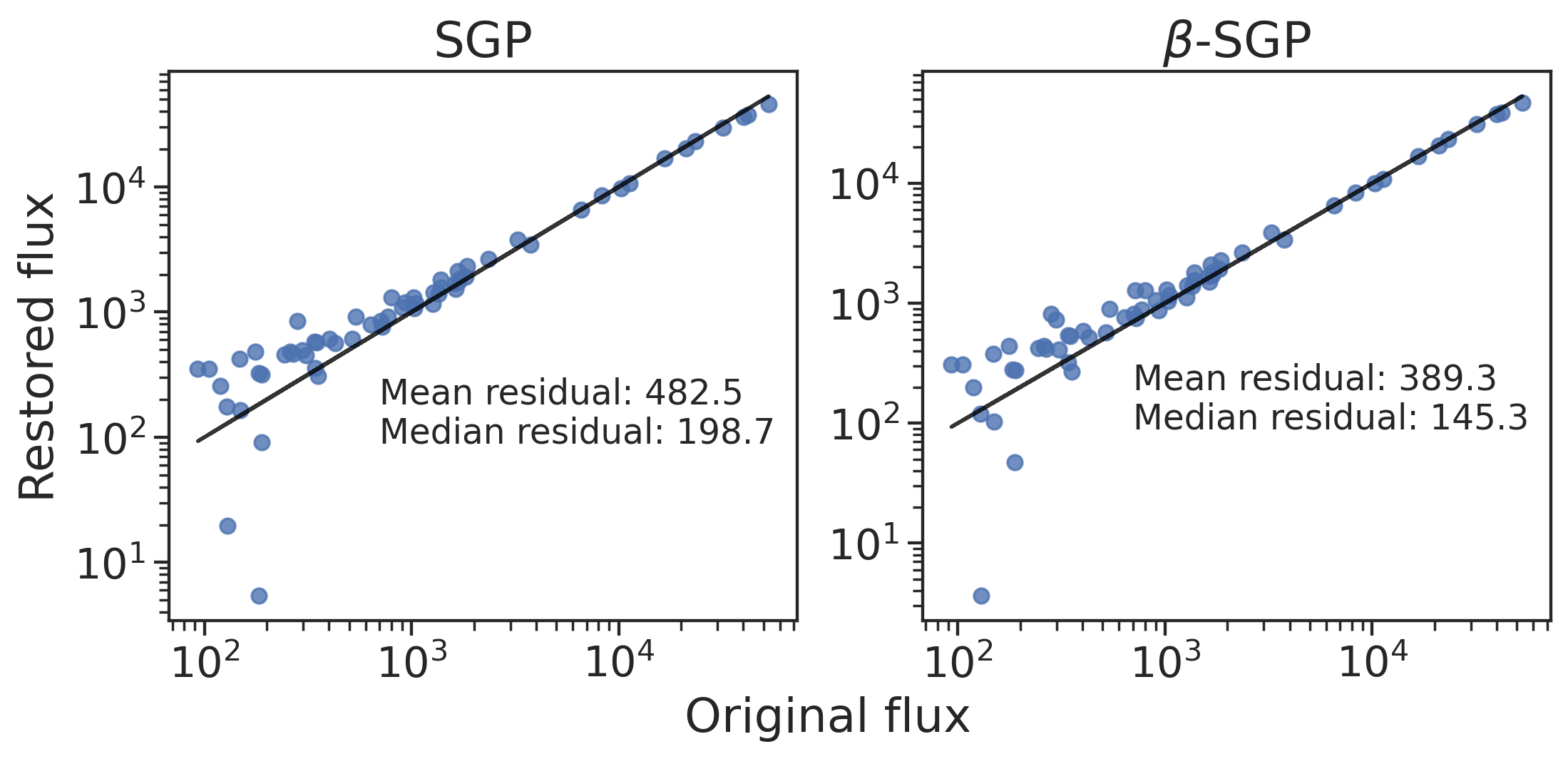}
        \caption{Star-by-star flux (in counts) comparison of matched sources from the original and restored subdivisions. Flux uncertainties are not shown for easier visualization. Both SGP and $\beta$-SGP generate 62 matches. The mean residual is the mean of the absolute value of differences in the original and restored flux values. Similarly, for the median residual. It quantifies the flux agreement between the original and restored images. $\beta$-SGP has a lower residual, which indicates it preserves the flux better than SGP.} \label{fig:ex2}
    \end{subfigure}%
    \caption{Example restoration of a subdivision extracted from the outskirts of the globular cluster (subdivision $1_5$: first along the x-direction and fifth along the y-direction) to prevent significant overlapping of stellar profiles.}
    \label{fig:subdivExample}
\end{figure}

For this experiment, we visually selected an image with distorted stars (having non-circular, highly elliptical, or other weird shapes) to compare the capabilities of SGP and $\beta$-SGP to restore distorted stellar shapes in our data. As described in Sect.~\ref{sec:obsData}, we generated $5 \times 5$ subdivisions (each of size $\sim400 \times 400$ pixels) from this image. For this experiment, we selected a subdivision towards the edge of the globular cluster, which is the sparsest field possible for our dataset. Fig.~\ref{fig:subdivExample} shows a comparison of the restored subdivisions using SGP and $\beta$-SGP.

Unlike the case in Sect.~\ref{sec:starStampsResults}, there are multiple sources in a single image, so we compare the original and restored flux source-by-source in the below panel (Fig.~\ref{fig:ex2}). Sources are detected, and their photometries are calculated using the \texttt{SourceCatalog} class as detailed in Sect.~\ref{sec:setup}. We use the same source detection parameters for all three images (e.g., 1.5$\sigma$ detection threshold, the FWHM of the Gaussian kernel used before detection = 1.2, and other background estimation parameters). We then performed crossmatching of sources from the original and the restored subdivisions using the TOPCAT software \citep{2005ASPC..347...29T}. We set a maximum permissible error of 1.1 pix while crossmatching so that any source within a 1.1 pix radius around the original source is considered a successful match. The symmetric, best-match criterion is used.

A visual comparison shows that the restored images are not drastically different. To quantify the stellar shapes, we calculated the median ellipticity and FWHM of the detected stars in the original and the restored images using SGP and $\beta$-SGP. It is important to note that the total flux is exactly\footnote{up to 3-4 decimal places} preserved for both SGP and $\beta$-SGP, which is a direct consequence of the flux-conserving constraint. However, it is of more interest to us to compare the source-by-source flux comparison. In terms of that, $\beta$-SGP has a lower mean and median residual and hence outperforms SGP. This indicates that $\beta$-SGP not only preserves the flux of stars better than SGP when considered individually (as seen in Sec.~\ref{sec:starStampsResults}) but also in an image containing multiple stars.

As a quick test, we also compared the flux residuals at the faint regime to see how the flux conservation compares for fainter sources. We visually selected a cutoff flux by looking at Fig.~\ref{fig:ex2}, which shows that for original flux values $\lesssim 2 \times 10^3$, the scatter increases. Hence, we calculated the mean and median residual only considering points with flux less than the set cutoff. We still observe better flux conservation using $\beta$-SGP (a reduction of $\sim$13 counts for the mean residual and $\sim$21 counts for the median residual). This shows that $\beta$-SGP also preserves the flux of faint sources better than SGP. A more extensive analysis comparing the flux of unmatched sources, and repeating this for a large set of images, will give an exact picture of the capabilities of $\beta$-SGP in recovering the faint ends better. 

The three marked positions, A, B, and C, require some discussion. Just to the right of A, a large blob was detected as a single source. Whereas, at the corresponding places in both the restored images, three sources closely separated were detected. This is an example of the deblending capacity of SGP and $\beta$-SGP. A similar observation can be made around point B, although, in this case, the source finder could detect two distinct sources in the original image itself. However, in the restored images, the presence of those two sources becomes apparent visually. Similarly, below point C, we see a new source is detected in the restored images just above the big bright source in the original image, which could hardly be seen in the original image. This discussion shows that both SGP and $\beta$-SGP are capable of deblending closely separated images. The observation in case C is particularly important since it shows that the restoration could deblend a faint source nearby a bright source.


\begin{table}
       \centering
       \begin{tabular}{ccccc}
        \hline
        Image & Median FWHM & Median Ellipticity\\
        \hline
        Original & 4.5 & 0.305\\
        SGP-restored & 2.42 & 0.154\\
        $\beta$-SGP-restored & 2.39 & 0.148\\
        \hline
       \end{tabular}
       \caption{Comparison of median FWHM and ellipticities of the original and restored stars from the subdivisions shown in Fig.~\ref{fig:ex1}. The median is taken from only the matched sources and not all detected sources in that image. $\beta$-SGP yields lower values than SGP, indicating an improvement in the restoration.}
       \label{table:subdivRes1}
   \end{table}




\subsubsection{Subdivision extracted from the center of the globular cluster}

Now, we perform restoration using both algorithms on a subdivision containing many closely spaced stars. To achieve this, we selected a subdivision from the center of the globular cluster. It serves as an extreme case, where not only are multiple stars present, but there is also a significant blending of their profiles.

\begin{figure}
    \centering
    \begin{subfigure}{\textwidth}
        \includegraphics[width=\textwidth,keepaspectratio]{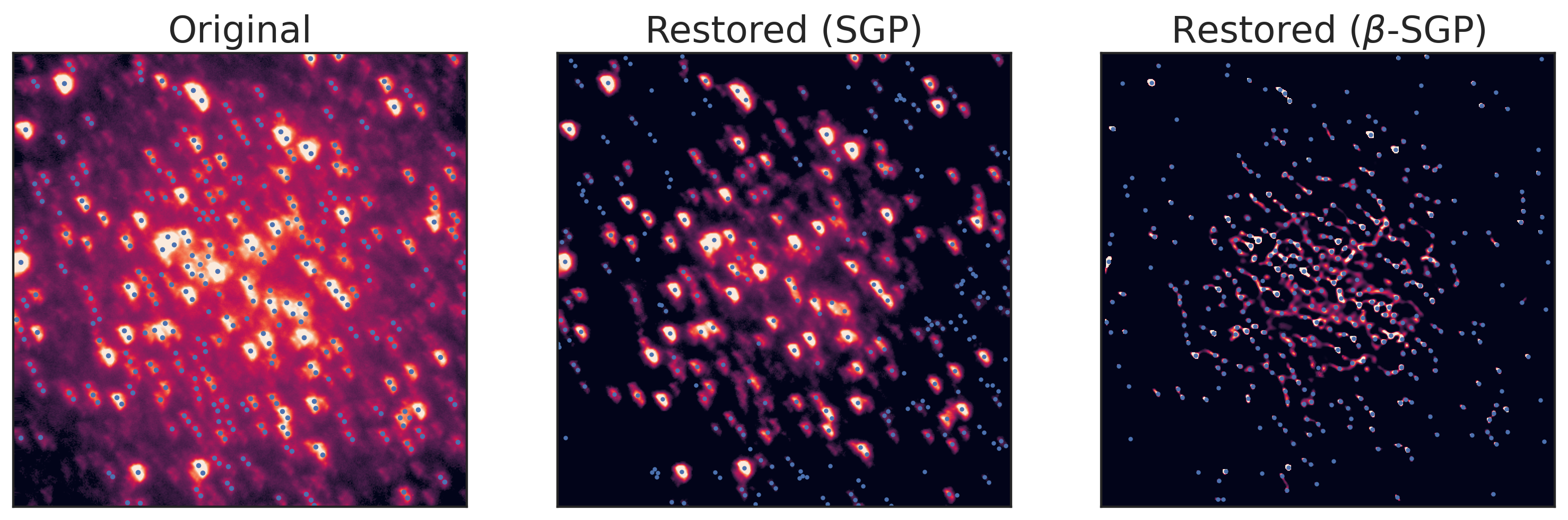}
        \caption{Detected sources are marked with a blue dot. The original image contains 392 detected sources, and the restored image using SGP and $\beta$-SGP contain 321 and 407 sources, respectively. SGP and $\beta$-SGP required 2 and 51 iterations, respectively, and execution time of $\sim$0.6s and $\sim$10.2s, respectively. The optimal $\beta_{init}$ for $\beta$-SGP found was $\sim$1.025. Images are shown using a combination of square root stretching and clipping values outside the central 95\% percentile pixel values for the original and 97\% for both the restored images.} \label{fig:crowded1}
    \end{subfigure}%
    \hfill
\begin{subfigure}{\textwidth}
        \includegraphics[width=\textwidth,keepaspectratio]{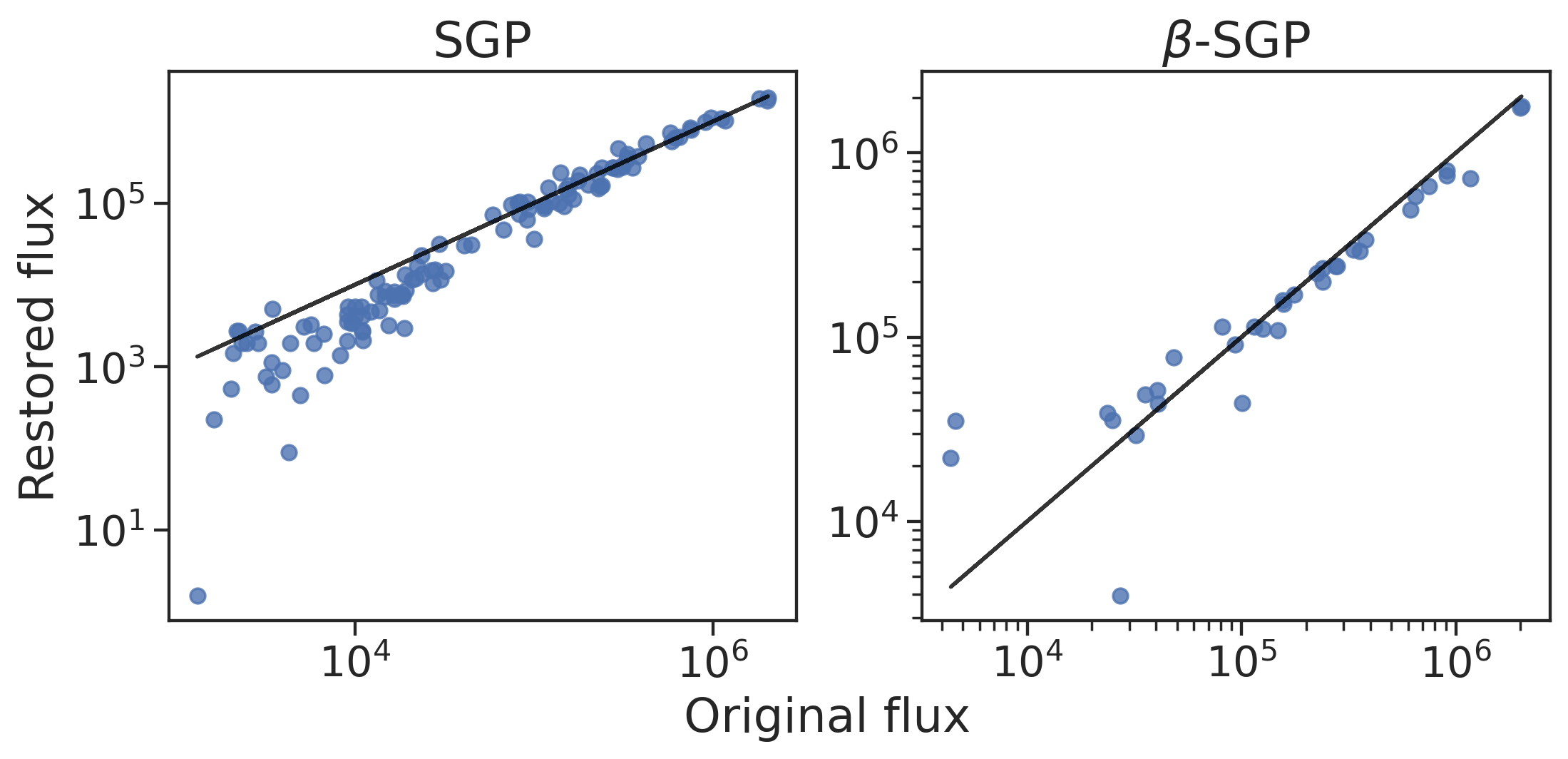}
        \caption{Star-by-star flux (in counts) comparison of matched sources from the original and restored subdivisions. SGP generated 121, and $\beta$-SGP generated 35 matches. Due to the vastly different sample sizes (discussed in the main text in detail), we do not show the mean and median flux residual.} \label{fig:crowded2}
    \end{subfigure}%
    \caption{Example restoration from a subdivision extracted from the center of the globular cluster (subdivision $3_3$: third along the x and y-directions, i.e., the middle subdivision), serving as an extreme case of overlapping.}
    \label{fig:crowdedExample}
\end{figure}

Fig.~\ref{fig:crowdedExample} shows the restoration results for this case. Unlike the previous sections, the restored images using SGP and $\beta$-SGP vastly differ visually. This is not unexpected since SGP only used two iterations whereas $\beta$-SGP used 51 iterations, as described in the caption of Fig.~\ref{fig:crowded1}\footnote{Our internal experiments have suggested that the number of iterations and the visual appearance of the restored image using SGP are sensitive to the choice of the detection threshold. While we found 1.5$\sigma$ to be a good choice for allowing the detection of fainter stars and hence including them for calculating the total flux of all stars in the image, SGP's restored image using a 2$\sigma$ threshold looks visually similar to $\beta$-SGP's restoration using 1.5$\sigma$ shown in the main text and takes much more than two iterations. However, such a detection threshold choice leaves out $\sim$30-35 faint sources. Hence, we do not show results with such a choice.}. Visually, one can observe that the restored image using $\beta$-SGP reconstructs a star profile closer to a circular shape and makes them more compact than the stars in the original image, which had a triangular-like shape. One might observe that some stars at the edges do not lie entirely in the original image; however, the restored images using both approaches still preserve and restore their shapes successfully.

Successfully and reliably cross-matching sources from crowded stellar fields or fields with worse seeing is an ongoing research problem (e.g., \citealt{2017MNRAS.468.2517W}; \citealt{2019ApJ...870...51S}). Instead of devising the best ways to crossmatch sources in our crowded and large-seeing field, we opt for simplicity and focus on the matches generated by TOPCAT\footnote{Another option for crossmatching would be crossmatching not with the (degraded) original image, but a good seeing image of the same field. However, we have not opted for that here since other issues can arise due to the variability of stars, such that a star is present in one image but not in the other. However, it can still be done for a fair estimate since a large fraction of stars in GCs are not variable.}. In terms of the number of matches, $\beta$-SGP has a far lesser number of matches than SGP even though it detects $\sim$90 sources more than SGP\footnote{It is known that some care is required for crossmatching in crowded fields: using a symmetric, best match criterion can give unsatisfactory results since the sources are so close to each other that unwanted matches could be generated. In such cases, asymmetric matching is generally preferred. However, we have found no difference between symmetric and asymmetric matching for this crowded field, due to which we stick with symmetric matching.}. The situation here is more complicated than in the previous experiments since the overlapping is so strong that almost a continuum of sources can be seen toward the center in the original image. $\beta$-SGP demonstrates the capability to resolve this crowded center of the globular cluster, due to which much more stars could be detected towards the center than were detected in the original image. However, the structure of this central core in the restored image using $\beta$-SGP has significantly changed. We attribute the apparent loss of matched sources using $\beta$-SGP to the enhanced resolution at the center and closely separated stars, which is worsened by the suppression of sources apart from the center. Due to this reason, the flux comparison, in this case, must be considered only a conservative comparison, i.e., using only those sources that were easy to match between the original and restored images. As in the previous section, the sum of the flux of all detected stars is exactly preserved, which is always going to be true as long as the flux-conserving constraint is used in the algorithm.

We also compare the properties of stars in the restored images in Table~\ref{table:subdivRes2}. As can be seen, $\beta$-SGP reduces the FWHM and ellipticity much more than SGP; it is expected based on the visual comparison of the restored images.

\begin{table}
       \centering
       \begin{tabular}{ccccc}
        \hline
        Image & Median FWHM & Median Ellipticity\\
        \hline
        Original & 4.69 & 0.307\\
        SGP-restored & 5.13 & 0.296\\
        $\beta$-SGP-restored & 2.52 & 0.239\\
        \hline
       \end{tabular}
       \caption{Comparison of median FWHM and ellipticities of the original and restored stars from the subdivisions shown in Fig.~\ref{fig:crowded1}. The median is taken from only the matched sources and not all detected sources in that image. $\beta$-SGP yields significantly lower values than SGP, indicating an improvement in the restoration. This can also be predicted visually by looking at the restored subdivisions. It must be noted that the sample sizes used for taking the median are quite different due to the inclusion of only matched sources. See main text for details.}
       \label{table:subdivRes2}
   \end{table}

Finally, we note that some relatively fainter stars towards the edges are suppressed for both approaches and can no longer be seen and detected, which is a limitation. As discussed before, this is very likely due to the projection step, due to which the brighter sources are given more preference. However, it can be potentially solved by experimenting with a smaller tolerance level or using smaller-sized subdivisions. Thus, with the current implementation, the faintest sources can be lost if there are many sources in the image and the difference in the fluxes (or magnitudes) between the brightest and the faintest objects in an image is high, i.e., the dynamical range is high. It is our expectation that in sparser fields with a high dynamical range, the faintest sources can still be recovered since there are not many sources so as to prefer some and suppress the others.

\section{Discussion}




In this paper, we performed single-image deconvolution where the PSF is assumed to be known, with a particular focus on the restoration of stellar shapes. This work was primarily inspired by the recent advancements in the applications of the SGP algorithm, which is theoretically and empirically proven to be more efficient than the famous RL deconvolution algorithm on a wide range of astronomical images such as nebulae, galaxies, and open star clusters. Due to its effective strategies for improved convergence, SGP has been recently studied extensively to test its plausibility as a possible replacement for RL.

We have showcased the capabilities of SGP to restore stars from real astronomical images, which is an important aspect since the application of SGP on non-simulated images has been through less scientific scrutiny. The aim of this paper is also to introduce $\beta$-divergence, a broader class of divergence measure that encompasses KL divergence as a special case in SGP. It is to be noted that in practice, these algorithms can be terminated using various criteria, and KL divergence is one of the commonly-used ones. Hence, the use of $\beta$-divergence makes sense when some divergence measure, like the KL divergence, is used as the stopping criterion. $\beta$-SGP, as we called it in this paper, is a generalized version of SGP without modifying the specifics of the SGP algorithm otherwise. We allowed automatic updates to the parameter, $\beta$, regulated by a simple stochastic gradient procedure. The learning rate for this update also changes by a specified rule in our formulation. We have found no concrete rule to set the initial value of $\beta$. Instead, we recommend selecting a few (e.g., $\sim$5-10) different values and selecting an optimal starting point based on some metric. Here we used flux conservation for selecting the optimal starting point, but it can be extended to any other metric or combination of metrics.

The gradient descent may not automatically proceed towards the optimal $\beta$ due to the complicated nature of the $\beta$-divergence loss landscape; hence, the dependence of the restoration result on the initial value of $\beta$. However, our experiments, which used a non-extensive optimization of the initial $\beta$ value (only five different trials of initial $\beta$), have suggested that $\beta$-divergence does provide some benefits owing to its flexible nature, despite the fact that for values of $\beta$ outside the range $[1, 2]$, the convexity of $\beta$ does not hold. $\beta$ values outside the convexity interval have also been empirically helpful in other contexts (e.g., the $\beta$-NMF multiplicative updates; see \citealt{fevotte}). The theoretical convergence results of $\beta$-divergence in SGP are not strictly dealt with in this paper, but we have shown that it works well in practice in a wide range of scenarios. Further improvements could be expected if more extensive tuning is performed. Our modifications are heuristic, and extensive empirical analyses are presented. We note that the generalization of gradient projection methods in general, although with different modifications, was dealt in \citet{sgpGeneralizedProjs}, so our work is in line with the generalization ideology mentioned there.

We first compared $\beta$-SGP with SGP using two simulated cases. Optimizing the initial value of $\beta$ using the relative pixel-wise error metric, we could find a suitable starting point that yields a lower error than the traditional SGP. While most previous works test the applications of SGP on simulated images where the noise and blurring conditions are fully controlled, we also provide extensive results on real astronomical images, ranging from individual star stamps to images containing multiple stars in a single frame. In the latter category, we use a massively crowded field (center of a globular cluster) and a relatively sparser field as our guiding examples. To quantify the restoration quality, we used the FWHM, ellipticity, and flux as our metrics. Our experiments on star stamps indicated that $\beta$-SGP shows a significant improvement in flux conservation compared to SGP while still keeping similar FWHM and ellipticities of the restored stars compared to SGP, sometimes even showing an improvement compared to SGP. On subdivisions containing multiple stars, $\beta$-SGP not only preserved flux better than SGP but also yielded smaller FWHM and ellipticity of the restored stars compared to SGP. In particular, by crossmatching stars from the restored and original subdivisions and comparing their FWHMs, ellipticities, and fluxes, we observed $\beta$-SGP shows benefits over SGP on both types of fields: the relatively sparser field and extremely crowded field (center of the globular cluster). This is quantified by a smaller median FWHM and ellipticity in the restored images using $\beta$-SGP than SGP. The comparison of flux conservation in the crowded fields was tricky due to the enhanced resolution obtained by the restored image using $\beta$-SGP. As a result, we could only make a conservative comparison of fluxes of matched sources. Overall, these experiments provide a strong indication that there could exist a $\beta \neq 1$ that can allow for better flux conservation and improve the quality of the restored images. Using a few visual examples, we discussed how deconvolution/restoration using SGP and $\beta$-SGP could be used for deblending, which is an important application of deconvolution methods in general.


The current limitation of SGP and $\beta$-SGP is that the flux of individual sources is only guaranteed to agree up to $\sim$1-5\% for sufficiently bright sources (flux $\gtrsim 10^3-10^4$), despite the fact that the total flux is always preserved. This is expected since SGP's constraints involve only total flux conservation. To improve on this aspect, a natural solution is to run the algorithm on smaller-sized subdivisions such that a good compromise between the number of sources present in the subdivision (it should be minimized for better source-by-source flux conservation) and the overall computational time to restore all subdivisions is made. Some research could be conducted so that the flux constraint is modified such that the flux of each source is preserved rather than the total flux; however, it could make the projection step more time-consuming since the nature of the flux constraint would have changed. Another point to note is that relatively fainter stars tend to get suppressed when using SGP and $\beta$-SGP, which is likely due to the simple nature of the constraints (non-negativity and flux conservation). However, we have shown a visual example where a fainter star around a bright star was not suppressed but was newly detected in the restored images. Thus we conclude that suppression of fainter stars is likely to happen when considering a bigger-sized image containing multiple sources. Using smaller-sized subdivisions could again potentially solve this issue. Along similar lines, one use of these deconvolution algorithms is to find fainter stars submerged in the background noise in the observed images. Hence, the unmatched sources during crossmatching can be inspected to see the typical fluxes of sources present in the deconvolved but not in the original -- supposedly, a large fraction of such sources will be the fainter ones, provided those fainter sources are not strongly suppressed.

In terms of computation time, $\beta$-SGP does require, on average more time per iteration than SGP since updates to $\beta$ are allowed -- the majority of the extra time $\beta$-SGP takes goes into calculating gradients of $\beta$-divergence w.r.t $\beta$. On star stamps (sized $40 \times 40$ pixels), we found $\beta$-SGP takes $\sim 1-2 \times 10^{-4}$ seconds more time per iteration than SGP. The difference rises to $\sim$0.03s per iteration for larger-sized subdivions containing multiple sources. $\beta$-SGP's execution time could be reduced by using GPU (for fast gradient calculations) and/or using tools such as JAX \citep{2021ascl.soft11002B} to speed up computation. If $\beta$ updates are not required for a particular case, a similar execution time as SGP is expected without using any external tools.

It is still an open question whether $\beta$ updates are necessary for improved performance; however, empirically, we have found it slightly beneficial. In our internal tests, we observed the same results for the crowded field (in terms of flux conservation, FWHM, and ellipticity). This indicates that $\beta$ update did not occur - that can happen when no updates to $\lambda_{k}$ occur - see Algorithm~\ref{SGPalgo}. However, we observed slightly worse results (in terms of FWHM and ellipticity) and almost similar flux conservation for the sparser field when not using $\beta$ updates.




For large-scale applications of the $\beta$-SGP approach, it could be an interesting line of research to learn $\beta$ on a training dataset and make it adapt to the test dataset with only a few iterations or keep it fixed after learning. No observed image has a perfect seeing, due to which such algorithms can improve the science extracted from any observed image. However, these algorithms are most important to the most severely degraded images, in which case an appropriate quantification of the seeing quality, or equivalently, the PSF shapes could be made (some efforts along this direction have taken place, e.g., \citealt{2018MNRAS.478.5671W}). Here, we restricted it to visual analysis, but appropriate quantification and automation for large-scale applications will become important.

\section{Conclusion}

\noindent The enhanced flexibility due to the incorporation of $\beta$-divergence (parametrized by the parameter $\beta$) in SGP has resulted in:
\begin{enumerate}
    \item An improvement in the flux conservation of stellar sources, which holds even at the faint ends of the original flux.
    \item A consistent improvement in the restoration quality in terms of stellar shapes (quantified using FWHM and ellipticity).
\end{enumerate}
Despite the strong theoretical reasoning to use KL divergence for images corrupted with Poisson noise, we noted that the Poisson noise model is only a ``good" approximation made in astronomy, which in practice is not entirely true. Hence, it is natural to think that a $\beta \neq 1$ ($\beta = 1$ corresponds to KL divergence) could ``fit" the data better; however, large deviations from $\beta = 1$ are not likely to be helpful as studied in the literature previously.

Thus, generalizing the SGP algorithm using the $\beta$-divergence shows a promising alternative for improving the capabilities of the traditional SGP algorithm. In general, our study shows that using flexible and robust divergence measures could be an interesting line of research to improve current image restoration pipelines and are worth exploring. As such studies continue to grow, we will get a much better understanding of the capabilities and improvements availed by such generalized approaches.


This paper is focused on the introduction, testing, and analysis of the proposed restoration algorithm. The application of the algorithm for the restoration of the entire time-series globular cluster image dataset and image analysis post-restoration is part of future work, and the results will be reported in a separate communication.

\section{Code and data availability}
The codes associated with this manuscript are available at \url{https://github.com/Yash-10/beta-sgp/}. We also make the data (consisting of M13 globular cluster processed images used in this study) publicly available\footnote{\url{https://drive.google.com/file/d/13Vk2TpXgSB6IoLUIv-zdh-XI53wJp-0y/view?usp=sharing}}. 

\section{Acknowledgements}

This research made use of Photutils, an Astropy package for detection and photometry of astronomical sources \citep{astropyPhotutils}. We thank PRISMA, a project of the Italian Ministry of University and Research, grant 2008T5KA4L, for making their SGP code publicly available. M.S. acknowledges the financial support by the Department of Science and Technology (DST), Government of India, under the Women Scientist Scheme (PH), project reference number SR/WOS-A/PM-17/2019. Y.G. thanks Ashish Mahabal of Caltech for insightful discussions. S.S. thanks Archana Mathur of Nitte Meenakshi Institute of Technology and Management, Bangalore, India, for the discussions.

\appendix
\section{PSF modeling: mathematical details}\label{appen-psfMatCalc}
\begin{figure*}
\includegraphics[width=\linewidth]{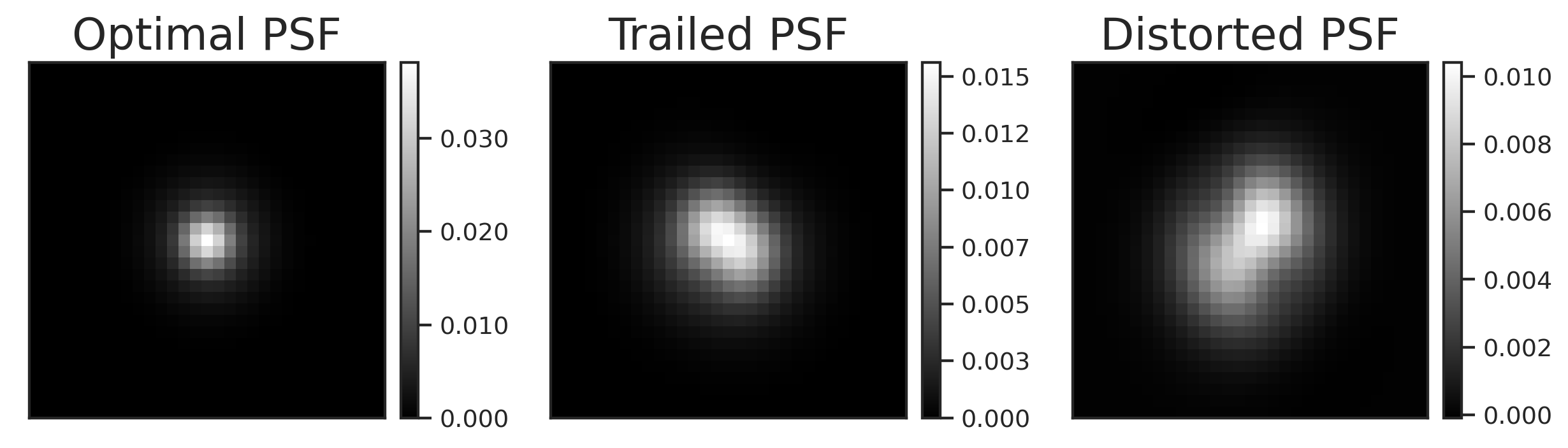}
\caption{Examples of three categories of PSF matrices observed in our dataset: (a) Optimal (a near-circular shape that best depicts the true star shape), (b) Trailed (an elliptical shape with elongation in one particular direction), and (c) Distorted (weird shapes that cannot be classified as either near-circular or elliptical shapes) obtained using the PSF model parameters from the DIAPL package.} \label{fig:1}
\end{figure*}

In summary, DIAPL's procedure to calculate the PSF model parameters is as follows: First, it reads the stellar coordinates output from the \codeword{sfind} program along with a bad pixel mask file. Then it rejects stars that lie close to each other or near the edges. It also performs an isolation test for the stars. For this, it only selects stars that are above the \textrm{NSIG\_DETECT}$\times \sigma$ level and rejects any star whose flux is contaminated due to crowding. We set \textrm{NSIG\_DETECT} to 2. On these well-isolated candidate stars, an initial symmetric circular PSF fit is made for a few iterations, followed by a final fit for another few iterations. The candidate stars might still be affected by cosmic rays or the blending of profiles with nearby stars. Hence, during the final fit, a sigma clipping procedure on the light distribution of the candidate star is performed to mitigate such undesirable scenarios. All of the parameters required during the whole process are user-defined. The \codeword{getpsf} routine calculates PSF vector coefficients used to build a PSF model in the form of a two-dimensional matrix. The size We produce PSF matrices of size $31 \times 31$ pixels. For this, we need to calculate the pixel value at each location, $(x, y)$, of the defined region, where $x, y \in [-hw, +hw]$, and $hw$ denotes the half-width of the PSF model, set to 15 for our case, resulting in a $31 \times 31$-pixels PSF model. In terms of programming, we start with a zero pixel value for each location in the raster and keep on accumulating values in a pixel based on the \textrm{NGAUSS} and \textrm{NDEG\_LOCAL} parameters, denoting the number of Gaussians used to build the PSF model (two in our case), and the degree of the polynomial used to describe the PSF shape, in our case set to two. The PSF model is a sum of two Gaussians, with the first Gaussian describing the core and the second describing the wings of the star. The second Gaussian is set to be $\approx 0.548$ times wider than the first, which is specified by the \textrm{SIGMA\_INC} parameter. It is beneficial to model both of them separately since both have different shapes and statistics; for example, the wings of any star are buried in photon noise, and the distance at which this happens is different for fainter and for brighter stars. After setting all the parameters, we reproduced DIAPL's \codeword{psf_core} script from the \codeword{phot} program in Python to calculate the values of the entries of the PSF matrix. Fig.~\ref{fig:1} shows the PSF matrices generated by this procedure. As mentioned in the main text, we generate PSF coefficients and, thus, the PSF matrix for each subframe separately without accounting for spatial variations.

\section{SGP and \texorpdfstring{$\beta$}{beta}-SGP: algorithm details}\label{appn:sgp-details}

The SGP algorithm, with a modification to update $\beta$, is shown in Algorithm~\ref{SGPalgo}, and $\beta$ update step is shown in line~13. As seen in line~15, we also schedule the learning rate using an exponential decay schedule given by
\begin{equation}
    \eta_i = \eta_0 \cdot e^{-k}\,,
\end{equation}
where $\eta_0$ is the initial learning rate, $k$ is the exponential decay parameter, and $\eta_i$ the learning rate at any iteration, $i \geq 1$.

We reproduce a Python implementation of the MATLAB code of SGP\footnote{\label{note1} \url{http://www.unife.it/prin/software}} for single-image deconvolution proposed by \citet{2012A&A...539A.133P}. We have modified the flux-conservation step to handle pixel saturation during the projection step alongside ensuring non-negativity of pixels: denoting the pixel counts above which the pixel is said to be saturated by $c$; we set the condition: $x_{i} = \min(x_{i}, c)$ so that we ensure no pixel is deemed saturated in the restored image. There are several reasons we incorporate this condition. First, if we do not account for saturation, photometry on the restored image would flag the star as saturated and exclude them from further analyses, which is undesirable. Second, after restoration, the same flux is spread among a smaller number of pixels (since the restored star would have a lower FWHM than the original star). Hence it is likely that a pixel exceeds the saturating condition if the pixels in the original image were already near saturation. Compared to the original SGP implementation, we have also allowed passing a two-dimensional background level, which could be helpful and yield better results for images with a strong gradient in the background across the field of view. Our tests of the original MATLAB SGP code showed that it could only work when the dimensions of the PSF model were the same as that of the image. As a result, we have optionally allowed using a two-dimensional Fast Fourier Transform-based convolution. When such an option is used, the difference is that we perform convolution by reshaping the image to two dimensions and then performing the convolution. In contrast, the original implementation performed it on the flattened (one-dimensional) images. For all our experiments on simulated images (in Sect.~\ref{sec:simResults}), we use the original \texttt{SGP-dec} code's implementation, whereas we use the two-dimensional convolution for all other experiments.

Choices for updating the scaling matrix include a diagonal matrix that approximates the inverse of the Hessian matrix, $\nabla^2 J(\bm{x})$ or a diagonal matrix that can be used to rewrite the RL method, in which the latter is computationally less expensive \citep{2009InvPr..25a5002B}. We use the latter option where the updating rule becomes,
\begin{equation}
    d_{i}^{k} = \textrm{min} \{L_{2}, \textrm{max} \{L_{1}, x_{i}^{k} \} \}\,,
\end{equation}
where $L_{1}$ and $L_{2}$ are the lower and upper bounds, respectively, on the elements of the scaling matrix. We use the appealing choice for the bounds that adapt themselves to the data described in \citet{2012A&A...539A.133P}.

As far as the projection step for flux conservation is concerned, we must solve a non-negative and linearly constrained strictly convex quadratic programming problem for flux conservation. Several linear-time projection algorithms exist in the literature (see \citealt{2009InvPr..25a5002B} for references). We use the secant-based approach suggested by \citet{dai-fletcher} that has shown good performance. For updating the step length parameter, we alternate between the two Barzilai \& Borwein step length (BB) rules \citep{barzBorw} only after the first 20 iterations, as suggested in \citet{2012A&A...539A.133P} (see Sec.~3 in \citealt{2009InvPr..25a5002B} for more discussion). This effective strategy also makes the choice of initial $\tau$ less important for convergence \citep{2009InvPr..25a5002B}. In step~3 of Algorithm~\ref{SGPalgo}, the subscript $+$ in the projection operator $P$ denotes the closed convex set containing $x$ that satisfies the flux conservation and the non-negativity constraints on $x$.

The parameters common to SGP and $\beta$-SGP are set as follows: $\beta_{b} = 0.4$, $\gamma = 10^{-4}$, $M = 1$, $\alpha_{min} = 10^{-5}$, $\alpha_{max} = 10^{5}$, $M_{\alpha} = 3$, $\tau = 0.5$, $\alpha_{0} = 1.3$, where all values are taken from \citet{2012A&A...539A.133P}, and use a maximum of 1000 flux conservation projections. Moreover, $M = 1$ implies that the line-search strategy reduces to the standard monotone Armijo rule \citep{2009InvPr..25a5002B}, and we have used it since we did not find any significant improvements by using a non-monotone strategy. Hyperparameters due to the inclusion of beta divergence are set as follows: initial learning rate for updating $\beta$, $\eta$ = $10^{-3}$, exponential learning rate schedule parameter, $k = 0.1$. The procedure we used to set the initial value of $\beta$ is described at various places throughout the text.

\begin{algorithm}
\caption{$\beta$-SGP}
\label{SGPalgo}
Choose the starting point $x^{(0)} \in \Omega$, and select $(\beta_{b}, \gamma) \in (0, 1)$; $\beta_{b}$: backtracking line search parameter, $\gamma$: line search penalty parameter, $0 < \alpha_{min} < \alpha_{max}$, $\alpha_{0}$: initial step length, $M_{\alpha}$: memory length, $\tau$: alternating parameter, memory length, $M > 0$, and the initial scaling matrix, $D_{0}$. Set lower and upper bounds for $D$. Set initial value of $\beta$ and learning rate, $\eta_0$, for updating $\beta$. $tol$ is the tolerance level.\\
\While {true}{
Projection: $\bm{y}^{(k)} = P_{+}(\bm{x}^{(k)} - \alpha_{k} D_{k} \nabla f(\bm{x}^{(k)})) $\\
\If{$\bm{y}^{(k)} = \bm{x}^{(k)}$}{break}
Descent direction: $\bm{d^{(k)}} = \bm{y}^{(k)} - \bm{x}^{(k)}$\\
Set $\lambda_{k} = 1$ and $f_{max} = \max_{0 \leq j \leq min(k, M-1)}  f(\bm{x}^{(k-j)})$\\
\While{true}{
\If{$f(\bm{x}^{(k)} + \lambda_{k} \bm{d^{(k)}}) \leq f_{max} + \gamma \lambda_{k} \nabla f(\bm{x}^{(k)})^T \bm{d^{(k)}} $ or $\lambda_{k} < tol$}{break}
\Else{$\lambda_{k} = \beta_{b} \lambda_{k}$ \\ $\beta = \beta - \eta \cdot \dfrac{1}{n} \sum \nabla_{\beta} f(\bm{x}^{k})$}
}
Set $\bm{x}^{(k+1)} = \bm{x}^{k} + \lambda_{k} \bm{d}^{k}$\\
Update the scaling matrix: $D^{(k+1)} \leftarrow D^{k}$, the step length: $\alpha^{(k+1)} \leftarrow \alpha^{k}$, and the learning rate, $\eta$.\\
}
\end{algorithm}



\end{document}